# Ferromagnetic Mott-Insulating State in Double Perovskites $Gd_2MgIrO_6$

Madhav Prasad Ghimire[1*], and R. K. Thapa[1,2]

[1]Condensed Matter Physics Research Center, Butwal, Rupandehi, Nepal
[2]Department of Physics, Mizoram University, Aizawl 796009, Mizoram, India

**Abstract.** We have studied the electronic and magnetic properties of double perovskites $Gd_2MgIrO_6$ by first-principles density functional theory (DFT). Based on the DFT calculations, $Gd_2MgIrO_6$ is found to have a ferromagnetic (FM) ground state. The material undergo half-metallic ferromagnets to Mott-Hubbard insulator transition which happens due to strong correlation in Gd-4f and Ir-5d states. Our results shows that the 5d electrons of Ir hybridize strongly with O-2p states near the Fermi level giving rise to the insulating state of $Gd_2MgIrO_6$. Our study suggests that the enhanced magnetic moment is a result of itinerant exchange rather than the exchange interaction involving individual ions of Gd and Ir atoms. The total magnetic moment calculated in the present studies is 15 $\mu_B$ per formula unit for $Gd_2MgIrO_6$.

**Keywords:** First-principles calculations, strong-correlation effects, ferromagnetism, metal-insulator-transition, Mott-Hubbard insulator.

**PACS:** 71.15.Mb; 71.27.+a; 71.20.Eh; 75.50.Cc; 71.30.+h.

## INTRODUCTION

Double perovskite oxides with general formula $A_2BB'O_6$, where A being an alkaline earth metal such as Sr, La, Gd, etc. and B and B' sites are occupied alternately by different transition metals, MgIr, LiIr, FeMo, FeRe, MnMo, etc [1-2]. In these class of compounds, the super exchange interactions between B and B' ions are in the form of B–O–B', instead of B–O–B form in simple perovskites of $ABO_3$. Perovskites oxides have been studied since 1950's when ferromagnetic behavior is observed in manganites (A=divalent or trivalent cations) around room temperature by Jonker and Santen [3]. In these compounds, the existence of mixed valence in Mn allowing for an electron transfer through oxygen orbitals was invoked in order to explain the ferromagnetic behaviour via a double exchange mechanism proposed by Zener [4]. Observation of half-metallic ferromagnetism (HMF) in Heusler alloys and double perovskite such as $Sr_2FeMoO_6$ triggered a new interest in these types of compounds when a considerable magneto-resistance behavior and a high magnetic transition temperature are observed which are important for the operation of devices at room temperatures [5-6]. Due to the novel properties such as spin-polarized current in absence of external field, high magnetic ordering temperature, giant magneto-resistance (GMR) and magneto-dielectricity, double perovskites are suitable candidates for spintronic device applications.

Experimentally there had been a number of study on crystal growth, crystal structure and magnetic properties of double perovskites oxides. Glasso and Darby [7] repored the existence of the compounds $La_2MIrO_6$ for M=Mg, Mn and Ni while Blasse [8] reported the occurrence of the isostructural compounds for M=Mg, Co, Ni and Cu. Currie et al. [9] also studied the structure and magnetic properties of $La_2MIrO_6$ (M=Mg, Co, Ni and Zn). The magnetic susceptibility measurement shows magnetic ordering for M=Co and Ni. Likewise, Mugavero et al. [10] studied a series of lanthanide oxides $Ln_2MgIrO_6$ (where Ln=Pr, Nd, Sm-Gd) and determine the ferromagnetic as well as anti-ferromagnetic behavior. Recently, the correlated metal oxides, especially iridates and rhodates, have attracted extensive interest that leads to unconventional phases [11-13]. For example, $Ln_2Ir_2O_7$ is predicted to transform from a topological band insulator to a topological Mott insulator [12]. A newly synthesized double perovskite $Gd_2MgIrO_6$ (GMIO) from the family of iridates is of particular interest [10] due to its unique properties: (i) A-site element Gd and B'-site element Ir provides charge as well as spin, and (ii) topmost occupied states close to Fermi level ($E_F$) are exclusively spin-down bands contributed by $d$ electrons of Ir atoms, and (iii) $IrO_6$ octahedron exhibit large crystal distortion which may induce strong crystal field that helps in splitting the spin-up and spin-down bands near $E_F$. Hence, in the present work efforts have been taken to study the electronic and magnetic properties of double perovskites $Gd_2MgIrO_6$ by first-principles density functional approach.

## CRYSTAL STRUCTURE AND METHODS

$Gd_2MgIrO_6$ crystallize in the space group $P2_1/n$ with the monoclinic-distorted double perovskites structure as shown in Fig. 1. The titled compound has structural distortions due to the tilting and rotation of $IrO_6$ octahedron in addition to the different bond lengths between Ir and oxygen atoms.

The electronic and magnetic properties of $Gd_2MgIrO_6$ are studied by using full-potential linearized augmented plane wave (FP-LAPW) method based on density functional theory (DFT) as implemented in the WIEN2k code [14]. The core states are treated fully relativistically while the semi-core and valence states as treated semi-relativistically. The standard generalized-gradient approximation (GGA) exchange correlation potential within the PBE-scheme were used with Coulomb interaction U[15]. The results shown here is with $U_{Gd}$=6 eV and $U_{Ir}$=1.25 eV respectively[11]. We have chosen the muffin-tin (MT) radii for Gd, Mg, Ir, O1, O2 and O3 to be 2.34, 1.92, 1.99, 1.76, 1.76 and 1.77 a.u. respectively. Integrations in reciprocal space were performed using 108 spatial $k$-points in the irreducible wedge of the Brillouin zone.

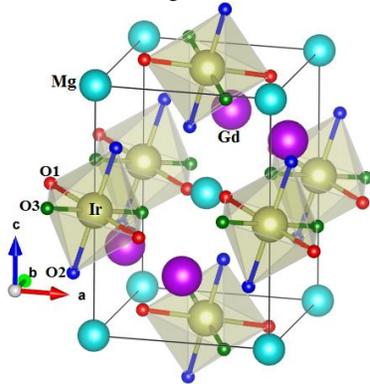

**FIGURE 1.** Crystal structure of the double perovskite $Gd_2MgIrO_6$. Distortion due to tilting and rotation of $IrO_6$ octahedron are shown.

## RESULTS AND DISCUSSIONS

GMIO belongs to double perovskite where Gd at A-site provides charge to the system and nominally take the charge state +3 with $4f^7$ configuration, lying deep in the valence region below the Fermi level ($E_F$). Unlike other cases however, Gd is in a high-spin state due to strong Hund's coupling. The transition element Ir nominally takes the charge state +4 with $5d^5$ configuration where five of the totally six $t_{2g}$ orbits are occupied and lie at the top of the valence band, forming a low-spin state due to large crystal field from oxygen octahedron.

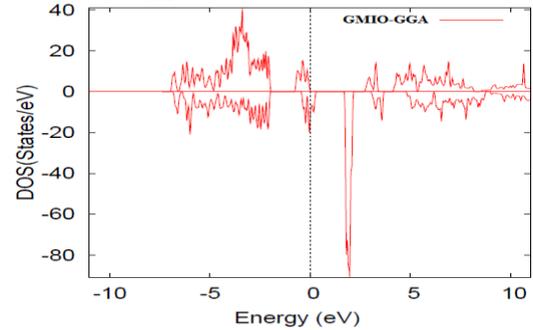

**FIGURE 2.** Total density of states (DOS) for $Gd_2MgIrO_6$ within GGA.

To know the ground state electronic and magnetic properties we calculated the total energy for nonmagnetic (NM), ferromagnetic (FM) and different antiferromagnetic (AFM) configurations. We observed that the FM configuration is the stable ground state consistent with the experimental prediction [9] with total energy of ~2 eV less than the NM and ~6 meV less than the AFM's configurations. According to the first-principles DFT calculations, the material GMIO shows half-metallic ferromagnetic behavior within GGA functional (see Fig. 2) with insulating state in spin-up and metallic state in spin-down channels. Since Gd and Ir are strongly correlated, GGA+U was considered wherein the material is found to be a Mott-Hubbard insulator with an energy gap of approx. 1.8 eV at $E_F$ (see Fig. 3 and Fig. 4).

From the total and partial (Gd-4f, Ir-5d, O-2p) DOS shown in Fig. 3 and the band structure in Fig. 4 for spin-up and spin-down channels, it is observed that the occupied Gd-$4f$ states lie deep in the valence region for spin-up channel whereas for spin-down, they appear far from $E_F$ in the conduction region. This gives rise to large exchange energy splitting (~6eV) in Gd. The Ir-$5d$ states are found to play key role in dictating the electronic properties of GMIO. They are found to hybridize strongly with the O-2p states in both spin channels. Their hybridization occurs mostly in the valence region near $E_F$ and in the conduction region. We observe that Ir-$t_{2g}$ (i. e., $d_{xy}$, $d_{xz}$, and $d_{yz}$) states are fully occupied in spin-up channels and thus lies in the valence region whereas in spin-down channel, only two out of three $t_{2g}$ states (i. e., $d_{xy}$, $d_{xz}$) are occupied while the remaining $d_{yz}$ states being empty lies in the conduction region along with the other un-occupied Ir-$de_g$ states. This picture is found consistent with the ionic picture. Charge transfer effect is prominent between Ir-5d and O-2p states due to strong hybridization. This induces sizable moment in oxygen atoms which get polarized in parallel with the Ir atoms. We observe that there is no exchange splitting between O-2p and Mg-2sp (not shown) electrons while for Gd-4f and Ir-5d DOS, there is an exchange splitting between spin-up and spin-down configurations. Exchange splitting energy of Ir-5d and Gd-4f is found to be ~2 eV and 6 eV respectively which contributes to ferromagnetic behavior in $Gd_2MgIrO_6$.

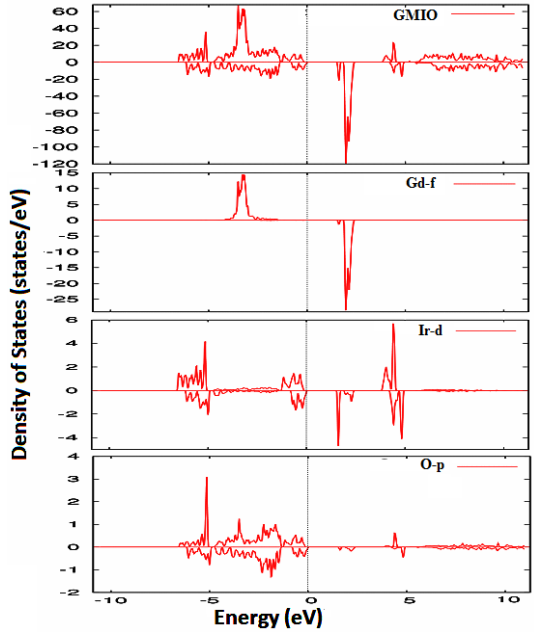

**FIGURE 3.** Total and partial DOS within GGA+U for $Gd_2MgIrO_6$: Gd-4f, Ir-5d and O-2p respectively.

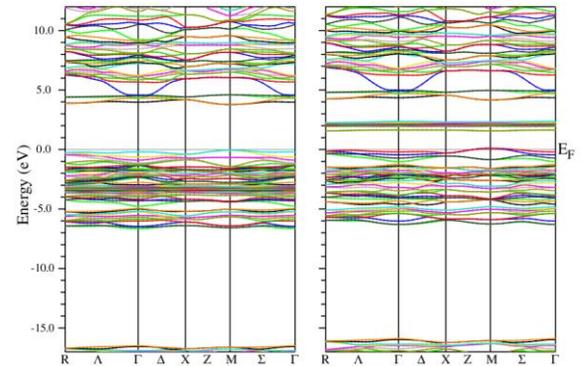

**FIGURE 4.** Band structures of $Gd_2MgIrO_6$ for spin-up (left) and spin-down (right) channels.

From first-principles calculations, we obtain magnetic moment of $\mu \approx +6.85\mu_B$ per Gd atom, $\mu \approx 0.71\mu_B$ per Ir atom and $\mu \approx 0.15\mu_B$ per one set of three O atoms, respectively, with a total angular moment of $\mu_{tot} = 15.03\mu_B$ per formula unit which is slightly larger than the experimental moment of $\mu_{tot} = 10.68\mu_B$ per formula unit. This discrepancy is expected to resolve when considering the spin-orbit coupling (work in progress) in Gd and Ir because both elements with half-filled shell will have their orbital moment anti-parallel to spin-moment and will reduce the effective moment close to the experimental moment.

## CONCLUSIONS

On the basis of first-principles density functional calculations, we propose $Gd_2MgIrO_6$ to be a ferromagnetic Mott-Hubbard insulator. Our study suggests that the enhanced magnetic moment is a result of itinerant exchange rather than the exchange interaction involving individual ions which can be resolved considering the spin-orbit coupling in Gd and Ir atoms.

## ACKNOWLEDGMENTS

MPG acknowledges the financial support from organizers of IWCCMP-2014 for the workshop. This work is supported partially by CMPRC, Butwal (Nepal) and the National plan for science, technology and innovation (India) under the research project No. 11-NAN1465-02.